\input harvmac

\Title{}
{\vbox{\centerline{Discrete Torsion and Branes in M-theory}
\vskip12pt\centerline{from Mathematical Viewpoint}}}

\centerline{Shigenori Seki\footnote{$^\dagger$}{seki@phys.h.kyoto-u.ac.jp} }
\bigskip
{\it \centerline{Graduate School of Human and Environmental Studies}
\centerline{Kyoto University, Kyoto 606-8501, Japan}}

\vskip .3in

\centerline{{\bf abstract}}

We study orbifold group actions on locally defined fields 
upon M-theory branes in a three-form C-fields 
background. We derive some constraints from the consistency 
of the orbifold group actions. We show the possibility 
of the existence of M-theory discrete torsion for the fields 
on the worldvolume and discuss its features.

\Date{March 2001}

\lref\SW{N.Seiberg and E.Witten, ``String Theory and Noncommutative Geometry'', JHEP 9909 (1999) 032, {\tt hep-th/9908142}.}
\lref\V{C.Vafa, ``Modular Invariance and Discrete Torsion on Orbifolds'', Nucl.Phys. B273 (1986) 592.}
\lref\VW{C.Vafa and E.Witten, ``On Orbifold with Discrete Torsion'', J.Geom.Phys. 15 (1995) 189, {\tt hep-th/9409188}.}
\lref\D{M.Douglas, ``D-branes and Discrete Torsion'', {\tt hep-th/9807235}.} 
\lref\DF{M.Douglas and B.Fiol, ``D-branes and Discrete Torsion II'', {\tt hep-th/9903031}.}
\lref\FHHP{B.Feng, A.Hanany, Y.He and N.Prezas, ``Discrete Torsion, Non-Abelian Orbifolds and the Schur Multiplier'', JHEP 0101 (2001) 033, {\tt hep-th/0010023}.}
\lref\FHHPii{B.Feng, A.Hanany, Y.He and N.Prezas, ``Discrete Torsion, Covering Groups and Quiver Diagrams'', {\tt hep-th/0011192}.}

\lref\SI{E.Sharpe, ``Discrete Torsion and Gerbes I'', {\tt hep-th/9909108}.} 
\lref\SII{E.Sharpe, ``Discrete Torsion and Gerbes II'', {\tt hep-th/9909120}.} 
\lref\SIII{E.Sharpe, ``Discrete Torsion'', {\tt hep-th/0008154}.} 
\lref\SIV{E.Sharpe, ``Analogues of Discrete Torsion for the M-theory Three-form'', {\tt hep-th/0008170}.} 
\lref\SV{E.Sharpe, ``Discrete Torsion in Perturbative Heterotic String Theory'', {\tt hep-th/0008184}.} 
\lref\SVI{E.Sharpe, ``Recent Developments in Discrete Torsion'', Phys.Lett. B498 (2001) 104, {\tt hep-th/0008191}.} 
\lref\SVIII{E.Sharpe, ``String Orbifolds and Quotient Stacks'', {\tt hep-th/0102211}.}
\lref\SIX{E.Sharpe, ``Quotient Stacks and String Orbifolds'', {\tt hep-th/0103008}.}
\lref\FW{D.Freed and E.Witten, ``Anomalies in String Theory with D-branes'', 
{\tt hep-th/9907189}.}
\lref\K{A.Kapustin, ``D-branes in a Topologically Nontrivial B-field'', Adv.Theor.Math.Phys. 4 (2001) 127, {\tt hep-th/9909089}.}
\lref\AOT{C.Ahn, K.Oh and R.Tatar, ``Orbifolds $AdS_7 \times S^4$ and Six Dimensional (0,1) SCFT'', Phys.Lett. B442 (1998) 109, {\tt hep-th/9804093}.}
\lref\FKPZ{S.Ferrara, A.Kehagias, H.Partouche and A.Zaffaroni, ``Membranes and Fivebranes with Lower Supersymmetry and their AdS Supergravity Duals'', Phys.Lett. B431 (1998) 42, {\tt hep-th/9803109}.}
\lref\BLI{D.Berenstein and R.Leigh, ``Discrete Torsion, AdS/CFT and Duality'', 
JHEP 0001 (2000) 038, {\tt hep-th/0001055}.}
\lref\BLII{D.Berenstein and R.Leigh, ``Non-Commutative Calabi-Yau Manifolds'', Phys.Lett. B499 (2001) 207, {\tt hep-th/0009209}.}
\lref\W{E.Witten, ``D-branes and K-theory'', JHEP 9812 (1998) 019, {\tt hep-th/9810188}.}
\lref\DMW{D.Diaconescu, G.Moore and E.Witten, ``$E_8$ Gauge Theory, and a Derivation of K-theory from M-theory'', {\tt hep-th/0005090}.}
\lref\DHOT{K.Dasgupta, S.Hyun, K.Oh and R.Tatar, ``Conifolds with Discrete Torsion and Noncommutativity'', JHEP 0009 (2000) 043, {\tt hep-th/0008091}.}
\lref\SS{S.Seki, work in progress.}

\newsec{Introduction}

NS-NS B-fields are studied in various contexts, particularly
noncommutative geometry and discrete torsion. 
The relation between the noncommutative geometry and the B-fields 
has been mentioned in \SW\ recently.
When non-trivial B-fields exist along a D-brane, 
a gauge theory on the D-brane has noncommutativity. 
On the other hand discrete torsion was originally pointed out in \V\ 
and the orbifold group action on a worldvolume gauge theory 
on D-branes at the orbifold 
singularity has been studied recently in \refs{\D,\DF} . 
In this direction a lot of works 
\refs{\FHHP\FHHPii\BLI\BLII{--}\DHOT} have been done, and 
this paper also has a great interest in this subject.

Since orbifold theories have twisted sectors,
constraints for modular invariance on a one-loop partition function of a 
closed string on an orbifold $M/\Gamma$, which is a quotient space of 
a manifold $M$ by an orbifold group $\Gamma$, induce a degree of freedom, 
which is called discrete torsion 
$\epsilon (g, h) \in U(1)$ for $g, h \in \Gamma$ \V. 
Higher loop modular invariance requires the following constraints 
\eqn\modinv{
\eqalign{
&\epsilon(g_1 g_2,g_3)=\epsilon(g_1,g_3)\epsilon(g_2,g_3) , \cr
&\epsilon(g,h) = \epsilon(h,g)^{-1} , \cr
&\epsilon(g,g) = 1, 
}}
where $\epsilon(g,h)$ is classified by a second group 
cohomology $H^2(\Gamma,U(1))$ .

Discrete torsion for open strings has been shown in \refs{\D, \DF}. 
For the orbifold $M/\Gamma$, supersymmetric Yang-Mills
fields $\phi$ in the worldvolume gauge theory 
on D-branes at the orbifold singularities 
are projected by $\Gamma$ as 
$$
\gamma(g)^{-1} \phi \gamma(g) = r(g) \phi, \quad g \in \Gamma, 
$$
where $\gamma(g)$ is a representation of $\Gamma$ in the gauge group 
and $r(g)$ is a space-time action of $\Gamma$. Then $\gamma(g)$ is in 
a projective representation
\eqn\d{
\gamma(g)\gamma(h) = \epsilon(g,h)\gamma(gh) .
}
The moduli space of the gauge theory has 
the same structure as the expectations based on \VW . 
More studies on non-abelian orbifolds with discrete torsion have been done 
in \refs{\FHHP,\FHHPii} in terms of Schur Multipliers.

String theories have physical and mathematical aspects. 
Mathematical understanding of discrete torsion has been proposed in 
\refs{\SI\SII\SIII\SIV\SV{--}\SVI}, which define the 
B-fields on each local patch
in terms of gerbes and state
that discrete torsion is the choice of orbifold group 
action on the B-fields. Since there exist gauge symmetries, 
it is in fact not sufficient to define 
the orbifold action only on the base space. 
It is necessary to choose the action on fields as well.
These remarks include more applications. For example
the choice of orbifold group action on vector fields gives rise to 
degrees of freedom which are known as orbifold Wilson lines.

In Section 2 we briefly review these mathematical aspects of discrete torsion 
and branes in the string theories. 
Section 3 is devoted for the calculations of orbifold group actions 
on fields in M-theory. 
In Section 4 we present summaries and conclusions 
with some discussions on problems left for further study.

\newsec{Review of discrete torsion and branes in string theories}

In this section we review the mathematical aspects of discrete torsion 
and branes proposed in \refs{\SI\SII{--}\SIII}. 
When we define the NS-NS B-fields, we need a two-form field $B^{\alpha}$ 
on a patch $U_{\alpha}$, a one-form $p^{\alpha\beta}$ 
on $U_{\alpha\beta} ( \equiv U_{\alpha} \cap U_{\beta}) $ and 
a $U(1)$-valued function $q_{\alpha\beta\gamma}$ on 
$U_{\alpha\beta\gamma} (\equiv U_{\alpha} \cap U_{\beta} \cap U_{\gamma})$
which satisfy the following equations 
\eqnn\bp
\eqnn\pq
\eqnn\qqq
$$\eqalignno{
B^{\alpha} - B^{\beta} &= d p^{\alpha\beta} , &\bp \cr
p^{\alpha\beta} + p^{\beta\gamma} + p^{\gamma\alpha} &= 
d \log q_{\alpha\beta\gamma} , &\pq \cr
\delta(q_{\alpha\beta\gamma}) &= 1 . &\qqq
}$$
From Eq.\qqq, in order to preserve {\v C}ech cocycle $q_{\alpha\beta\gamma}$ 
by an orbifold group action $g \in \Gamma$, up to coboundaries, 
we require that the pullback of $q_{\alpha\beta\gamma}$ becomes 
\eqn\gq{
g^* q_{\alpha\beta\gamma} = q_{\alpha\beta\gamma}
\nu_{\alpha\beta}^g \nu_{\beta\gamma}^g \nu_{\gamma\alpha}^g , 
}
where $\nu^{g}_{\alpha\beta}$ are some {\v C}ech cochains for each $g$.
From Eqs.\pq\ and \gq , we obtain the orbifold group action on 
$p^{\alpha\beta}$ as 
\eqn\gp{
g^* p^{\alpha\beta} = p^{\alpha\beta} + d \log \nu_{\alpha\beta}^g 
+ \Lambda(g)^{\alpha} - \Lambda(g)^{\beta} ,
}
for some one-forms $\Lambda(g)^{\alpha}$. Using Eqs.\bp\ and \gp , 
the orbifold group action on $B^{\alpha}$ becomes 
\eqn\gb{
g^* B^{\alpha} = B^{\alpha} + d \Lambda(g)^{\alpha} .
}
Expanding $(g_1 g_2)^* q_{\alpha\beta\gamma}$, 
$\nu_{\alpha\beta}^{g_1 g_2 g_3}$, $(g_1 g_2)^* p^{\alpha\beta}$ 
and $(g_1 g_2)^* B^{\alpha}$ respectively in two different ways, 
we obtain some constraints 
$$\eqalign{
\Lambda(g_1 g_2)^{\alpha} &= 
\Lambda(g_2)^{\alpha} + g_2^* \Lambda(g_1)^{\alpha} 
- d \log h_{\alpha}^{g_1,g_2} , \cr
\nu_{\alpha\beta}^{g_1 g_2} &= \nu_{\alpha\beta}^{g_2}
\left(g_2^* \nu_{\alpha\beta}^{g_1}\right)
\left(h_{\alpha}^{g_1, g_2}\right)
\left(h_{\beta}^{g_1, g_2}\right)^{-1} , \cr
\left(h_{\alpha}^{g_1,g_2 g_3}\right)
\left(h_{\alpha}^{g_2,g_3}\right) &= 
\left(g_3^* h_{\alpha}^{g_1,g_2}\right)
\left(h_{\alpha}^{g_1 g_2,g_3}\right) ,
}$$
where $h_{\alpha}^{g_1,g_2}$ are some {\v C}ech cochains.

Orbifold Wilson surfaces $\exp \left(\int B\right)$ which appear in 
the one-loop partition function, are the analogues 
of orbifold Wilson loops and give rise to phases. Now let us take
two kinds of definitions of the orbifold group actions. We describe
$h_{\alpha}^{g,h}$ in one definition and ${\bar h}_{\alpha}^{g,h}$ 
in the other. 
In order to consider the difference between these two definitions we use 
$$
\omega^{g,h} = {h^{g,h} \over {\bar h}^{g,h}}. 
$$
Then the phases lead to 
\eqn\stph{
\left(\omega^{g,h}\right)\left(\omega^{h,g}\right)^{-1}.
}
Note that we assume the B-fields are completely trivial so that 
$h^{g,h}_\alpha$ and $\omega_{\alpha}^{g,h}$ are globally defined.
Eq.\stph\ stands for the phases from the 
contribution of twisted sectors to the partition function, in other words, 
it corresponds to discrete torsion introduced in \V .
In fact the phases \stph\ satisfy the conditions \modinv\ 
for the modular invariance.

Next we consider the orbifold group action on $N$ 
coincident D-branes. There are $U(N)$ gauge fields, which come from the 
Chan-Paton factors of open strings ending on the D-branes. 
Since gauge transformations associate the 
gauge fields $A$ with the B-fields, the gauge fields also should be defined 
on local patches. From \FW\ we can link $A^{\alpha}$ to $B^{\alpha}$ 
by using the following equations 
\eqnn\gaugea
\eqnn\gaugeg
$$\eqalignno{
A^{\alpha} - g_{\alpha\beta} A^{\beta} g_{\alpha\beta}^{-1} 
- d \log g_{\alpha\beta}^{-1} &= p^{\alpha\beta} I , &\gaugea \cr
g_{\alpha\beta}g_{\beta\gamma}g_{\gamma\alpha} 
&= q_{\alpha\beta\gamma} I , &\gaugeg
}$$
where $g_{\alpha\beta}$ is a $N \times N$ matrix and $I$ is a unit matrix. 
$g_{\alpha\beta}$ is a transition function for the gauge bundle $A$ when 
the B-fields are completely trivial. 

From Eqs.\gq,\gp,\gaugea\ and \gaugeg\ 
the orbifold actions on the gauge fields become
\eqnn\ggaugea
\eqnn\ggaugeg
$$\eqalignno{
g^* A^{\alpha} &= (\gamma_{\alpha}^g) A^{\alpha} (\gamma_{\alpha}^g)^{-1} 
+ (\gamma_{\alpha}^g)d(\gamma_{\alpha}^g)^{-1} 
+ I \Lambda(g)^{\alpha} , &\ggaugea \cr
g^* g_{\alpha\beta} &= 
\left(\nu_{\alpha\beta}^g\right)
\left[(\gamma_{\alpha}^g)g_{\alpha\beta}(\gamma_{\beta}^g)^{-1}\right] ,
&\ggaugeg
}$$
for some $N \times N$ matrices $\gamma_{\alpha}^g$.
Expanding $(g_1 g_2)^* g_{\alpha\beta}$ in two different ways, 
we obtain a constraint 
\eqn\rvi{
(g_2^* \gamma_{\alpha}^{g_1})(\gamma_{\alpha}^{g_2}) =
h_{\alpha}^{g_1,g_2}(\gamma_{\alpha}^{g_1 g_2}) .
}%

Let us consider the completely trivial B-fields given by $B^{\alpha} = 0$, 
$p^{\alpha\beta} = 0$ and $q_{\alpha\beta\gamma} = 1$. 
Then it is meaningful that we set 
the bundle on the D-branes to be topologically trivial. We can replace
the locally defined gauge field $A^{\alpha}$ with a globally defined 
$U(N)$ gauge field $A$ and set $g_{\alpha\beta}=1$. 
If furthermore the gauge field $A$ is constant, 
we can assume that $\gamma^g$ and $h^{g_1,g_2}$ are constant.
From Eq.\rvi\ we obtain 
$$
\left(\gamma^{g_1}\right)\left(\gamma^{g_2}\right) 
= h^{g_1,g_2}\left(\gamma^{g_1 g_2}\right) .
$$
$\gamma^g$ has a projective representation.
And $h^{g_1,g_2}$ is classified by $H^2(\Gamma,U(1))$. 
These results are in good agreement with \refs{\D,\DF}.

\newsec{Discrete torsion and branes in M-theory}

In M-theory there exist membranes and M5-branes, and the membranes 
are considered as three-form C-fields in an eleven dimensional supergravity. 
Since the C-fields compactified
on $S^1$ lead to the B-fields in the string theories, 
membrane twisted sector would derive discrete torsion in M-theory.
In the string theories discrete torsion, 
which is classified by $H^2(\Gamma,U(1))$, 
appears as the phases derived from 
the term $\exp \left(\int B\right)$, while in M-theory the term 
$\exp \left(\int C\right)$ leads to some phases. In \SIV\ 
the phases have been calculated from  the contribution 
of membrane twisted sectors on $T^3$ in terms of 2-gerbes 
and have been classified by a third group cohomology $H^3(\Gamma,U(1))$.

In order to define the C-fields on each patch, 
we need a three-form C-field $C^{\alpha}$ on a patch $U_{\alpha}$, 
a two-form $u^{\alpha\beta}$ on $U_{\alpha\beta}$, 
a one-form $v^{\alpha\beta\gamma}$ on $U_{\alpha\beta\gamma}$ 
and a function $h_{\alpha\beta\gamma\delta}$ 
on $U_{\alpha\beta\gamma\delta} 
(\equiv U_{\alpha}\cap U_{\beta}\cap U_{\gamma}\cap U_{\delta})$.
The forms and the function are related by the following equations 
\eqnn\ci
\eqnn\cii
\eqnn\ciii
\eqnn\civ
$$\eqalignno{
C^{\alpha} - C^{\beta} &= d u^{\alpha\beta} , &\ci \cr
u^{\alpha \beta} + u^{\beta \gamma} + u^{\gamma \alpha}
&= d v^{\alpha \beta \gamma} , &\cii \cr
v^{\beta \gamma \delta} - v^{\alpha \gamma \delta} 
+ v^{\alpha \beta \delta} - v^{\alpha \beta \gamma}
&= d \log h_{\alpha\beta\gamma\delta} , &\ciii \cr
\delta h_{\alpha \beta \gamma \delta} &= 1 . &\civ
}$$
The actions of $g \in \Gamma$ for the C-fields described in \SIV\ are 
\eqnn\gci
\eqnn\gcii
\eqnn\gciii
\eqnn\gciv
$$\eqalignno{
g^* C^{\alpha} &= C^{\alpha} + d \Lambda^{(2)}(g)^{\alpha} , &\gci \cr
g^* u^{\alpha\beta} &= u^{\alpha\beta} + d \Lambda^{(1)}(g)^{\alpha\beta}
+ \Lambda^{(2)}(g)^{\alpha} - \Lambda^{(2)}(g)^{\beta} , &\gcii \cr
g^* v^{\alpha\beta\gamma} &= v^{\alpha\beta\gamma} 
+ \Lambda^{(1)}(g)^{\alpha\beta} + \Lambda^{(1)}(g)^{\beta\gamma} 
+ \Lambda^{(1)}(g)^{\gamma \alpha} \cr
&\phantom{ =\ } + d \log \nu^g_{\alpha\beta\gamma} , &\gciii \cr
g^* h_{\alpha\beta\gamma\delta} &= h_{\alpha\beta\gamma\delta} 
\left(\nu^g_{\beta\gamma\delta}\right) 
\left(\nu^g_{\alpha\gamma\delta}\right)^{-1}  
\left(\nu^g_{\alpha\beta\delta}\right)
\left(\nu^g_{\alpha\beta\gamma}\right)^{-1} , &\gciv
}$$
where, for each element $g$ in the orbifold group $\Gamma$, 
$\nu_{\alpha\beta\gamma}^g$ are some {\v C}ech cochains, 
$\Lambda^{(1)}(g)^{\alpha\beta}$ are some local one-forms and 
$\Lambda^{(2)}(g)^{\alpha}$ are some local two-forms. 
And we obtain the constraints \SIV ,
\eqnn\gcv
\eqnn\gcvi
\eqnn\gcvii
\eqnn\gcviii
\eqnn\gcix
\eqnn\gcx
$$\eqalignno{
\Lambda^{(2)}(g_1 g_2)^{\alpha} &= \Lambda^{(2)}(g_2)^{\alpha} 
+ g_2^* \Lambda^{(2)}(g_1)^{\alpha} + 
d \Lambda^{(3)}(g_1, g_2)^{\alpha} , &\gcv \cr
\Lambda^{(1)}(g_1 g_2)^{\alpha \beta} &= \Lambda^{(1)}(g_2)^{\alpha\beta} 
+ g_2^* \Lambda^{(1)}(g_1)^{\alpha\beta} - \Lambda^{(3)}(g_1,g_2)^{\alpha} \cr 
&\phantom{ =\ } + \Lambda^{(3)}(g_1,g_2)^{\beta} 
- d \log \lambda^{g_1, g_2}_{\alpha \beta} , &\gcvi \cr
\Lambda^{(3)}(g_2,g_3)^{\alpha} + \Lambda^{(3)}(g_1,g_2 g_3)^{\alpha}
&= g_3^* \Lambda^{(3)}(g_1,g_2)^{\alpha} + 
\Lambda^{(3)}(g_1 g_2,g_3)^{\alpha} \cr
&\phantom{ =\ } + d \log \gamma^{g_1,g_2,g_3}_{\alpha} , &\gcvii \cr
\nu^{g_1 g_2}_{\alpha\beta\gamma} &= 
\left(\nu^{g_2}_{\alpha\beta\gamma}\right)
\left(g_2^* \nu^{g_1}_{\alpha\beta\gamma}\right)
\left(\lambda^{g_1,g_2}_{\alpha\beta}\right)
\left(\lambda^{g_1,g_2}_{\beta\gamma}\right)
\left(\lambda^{g_1,g_2}_{\gamma\alpha}\right) , &\gcviii \cr
\left(\lambda^{g_1 g_2,g_3}_{\alpha \beta}\right) 
\left(g_3^* \lambda^{g_1, g_2}_{\alpha\beta}\right) &= 
\left(\lambda^{g_1,g_2 g_3}_{\alpha\beta}\right) 
\left(\lambda^{g_2,g_3}_{\alpha\beta}\right) 
\left(\gamma^{g_1,g_2,g_3}_{\alpha}\right) 
\left(\gamma^{g_1,g_2,g_3}_{\beta}\right)^{-1} , &\gcix \cr
\left(\gamma^{g_1,g_2,g_3 g_4}_{\alpha}\right)
\left(\gamma^{g_1 g_2,g_3,g_4}_{\alpha}\right) &= 
\left(\gamma^{g_1,g_2 g_3,g_4}_{\alpha}\right)
\left(\gamma^{g_2,g_3,g_4}_{\alpha}\right)
\left(g_4^* \gamma^{g_1, g_2, g_3}_{\alpha}\right) . &\gcx
}$$
In the similar way to Section 2 the difference of orbifold group actions 
$\omega_{\alpha}^{g_1,g_1,g_3} = 
\gamma_{\alpha}^{g_1,g_2,g_3}/{\bar \gamma}_{\alpha}^{g_1,g_2,g_3}$ leads
to the membrane twisted sector phase 
on $T^3$ and is classified by $H^3(\Gamma,U(1))$ 
for the completely trivial C-fields \SIV.

Now let us consider the orbifold group actions 
on fields in a worldvolume theory on branes. 
When a membrane ends on the branes as a string, 
the end line is assumed as a two-form 
field $B$ in the worldvolume theory. 
Since the three-form field $C$ is mapped to $C + dB$ by gauge transformations,
on the analogy of the string theories the B-fields are also to be 
defined on each local patch in terms of a two-form $B^\alpha$, a one-form 
$p^{\alpha\beta}$ and a function $q_{\alpha\beta\gamma}$. 
From \FW\ 
we are able to associate the B-fields with the C-fields as 
\eqnn\bi
\eqnn\bii
\eqnn\biii
$$\eqalignno{
B^{\alpha} - B^{\beta} + d p^{\alpha\beta} &= u^{\alpha\beta} , &\bi \cr
p^{\alpha\beta} + p^{\beta\gamma} + p^{\gamma\alpha} 
+ d \log q_{\alpha\beta\gamma} &= v^{\alpha\beta\gamma} , &\bii \cr 
\left(q_{\beta\gamma\delta}\right) \left(q_{\alpha\gamma\delta}\right)^{-1}
\left(q_{\alpha\beta\delta}\right) \left(q_{\alpha\beta\gamma}\right)^{-1} 
&= h_{\alpha\beta\gamma\delta} . &\biii
}$$
We calculate the orbifold group actions on the B-fields. Firstly we suppose 
the action on $q_{\alpha\beta\gamma}$ as 
$$
g^* q_{\alpha\beta\gamma} = 
q_{\alpha\beta\gamma} \left(\mu^g_{\alpha\beta}\right)
\left(\mu_{\beta\gamma}^g\right) \left(\mu^g_{\gamma\alpha}\right),
$$
so that $q_{\alpha\beta\gamma}$ is preserved up to coboundaries.
$\mu_{\alpha\beta}^g$ are some {\v C}ech cochains for each $g$.
Then the pullback of the left hand side of Eq.\biii\ becomes 
$$
g^*\left[\left(q_{\beta\gamma\delta}\right) 
\left(q_{\alpha\gamma\delta}\right)^{-1}
\left(q_{\alpha\beta\delta}\right) 
\left(q_{\alpha\beta\gamma}\right)^{-1}\right]
= \left(q_{\beta\gamma\delta}\right) 
\left(q_{\alpha\gamma\delta}\right)^{-1}
\left(q_{\alpha\beta\delta}\right) 
\left(q_{\alpha\beta\gamma}\right)^{-1} = h_{\alpha\beta\gamma\delta},
$$
while from Eq.\gciv\ the pullback of the right hand side of Eq.\biii\ 
has additional factors $\nu^g$. 
So we should instead define the orbifold action on $q_{\alpha\beta\gamma}$ as 
\eqn\gbiii{
g^* q_{\alpha\beta\gamma} = \left(\nu^g_{\alpha\beta\gamma}\right)
q_{\alpha\beta\gamma} \left(\mu^g_{\alpha\beta}\right)
\left(\mu_{\beta\gamma}^g\right) \left(\mu^g_{\gamma\alpha}\right). 
}
From Eqs.\gciii, \bii\ and \gbiii\ we obtain 
\eqn\gbii{
g^* p^{\alpha\beta} = p^{\alpha\beta} - d \log \mu^g_{\alpha\beta} 
+ \Lambda^{(1)}(g)^{\alpha\beta} + \Lambda^{(1)}(g)^{\alpha} 
- \Lambda^{(1)}(g)^{\beta} ,
}
for some local one-forms $\Lambda^{(1)}(g)^{\alpha}$. 
Using Eqs.\gcii, \bi\ and \gbii, we calculate the orbifold group action 
on $B^{\alpha}$ as 
\eqn\gbi{
g^* B^{\alpha} = B^{\alpha} - d \Lambda^{(1)}(g)^{\alpha} + 
\Lambda^{(2)}(g)^{\alpha} .
}
$\mu_{\alpha\beta}^g$ and $\Lambda^{(1)}(g)^{\alpha}$ determine the 
structure of orbifold group action on the B-fields.

If a membrane extends in a two dimensional subspace transverse to the branes
and ends on the branes as a point, we can 
consider the end point as a one-form field $A$ 
in the worldvolume gauge theory on the branes. 
From \FW\ we write down the relations between 
the B-fields and the A-fields,  
\eqnn\ai
\eqnn\aii
$$\eqalignno{
A^{\alpha} - g_{\alpha\beta} A^{\beta} g_{\alpha\beta}^{-1} 
- d \log g_{\alpha\beta} &= p^{\alpha\beta} , &\ai \cr
g_{\alpha\beta} g_{\beta\gamma} g_{\gamma\alpha} &= 
q_{\alpha\beta\gamma} . &\aii \cr
}$$
$A^{\alpha}$ and $g_{\alpha\beta}$ are the data of A-fields 
described on local patches in the same way as Section 2.
We suppose the orbifold group action on $g_{\alpha\beta}$ as 
\eqn\ggpro{
g^* g_{\alpha\beta} = 
\left(\rho_{\alpha}^g\right) g_{\alpha\beta} \left(\rho_{\beta}^g\right)^{-1}, 
}
where $\rho^g$ are some functions.
Since from the left hand side of Eq.\aii\ we obtain 
$$
g^* \left[g_{\alpha\beta} g_{\beta\gamma} g_{\gamma\alpha}\right] = 
g_{\alpha\beta} g_{\beta\gamma} g_{\gamma\alpha} = q_{\alpha\beta\gamma} ,
$$
we need additional factors in Eq.\ggpro\ in order for the above 
equation to be consistent with Eq.\gbiii .
When $\nu_{\alpha\beta\gamma}^g$ is equal to one, 
we are able to define the orbifold action on 
$g_{\alpha\beta}$ as
\eqn\gaii{
g^* g_{\alpha\beta} = \left(\mu^g_{\alpha\beta}\right)
\left(\rho^g_{\alpha}\right) g_{\alpha\beta} 
\left(\rho^g_{\beta}\right)^{-1} .
}
If $\Lambda^{(1)}(g)^{\alpha\beta}$ vanishes, 
we can obtain the action of $g$ on $A^{\alpha}$ as
\eqn\gai{
g^* A^{\alpha} = 
\left(\rho_{\alpha}^g\right) A^{\alpha} \left(\rho_{\alpha}^g\right)^{-1} 
+ d \log \rho^g_{\alpha} + \Lambda^{(1)}(g)^{\alpha} . 
}
These two requirements, $\nu_{\alpha\beta\gamma}^g = 1$ 
and $\Lambda^{(1)}(g)^{\alpha\beta} = 0$, are realized, for example, 
when the C-fields are topologically trivial.

Next we derive some constraints. We calculate 
$(g_1 g_2)^* q_{\alpha\beta\gamma}$ for $g_1, g_2 \in \Gamma$ 
in two different ways and they become
$$\eqalign{
g_2^* (g_1^* q_{\alpha\beta\gamma}) &=
\left(\nu_{\alpha\beta\gamma}^{g_2}\right) q_{\alpha\beta\gamma} 
\left(g_2^* \nu_{\alpha\beta\gamma}^{g_1}\right) \cr
&\phantom{=}\ \times \left(\mu_{\alpha\beta}^{g_2}\right)
\left(g_2^* \mu_{\alpha\beta}^{g_1}\right) 
\left(\mu_{\beta\gamma}^{g_2}\right)\left(g_2^* \mu_{\beta\gamma}^{g_1}\right) 
\left(\mu_{\gamma\alpha}^{g_2}\right)
\left(g_2^* \mu_{\gamma\alpha}^{g_1}\right) , \cr
(g_1 g_2)^* q_{\alpha\beta\gamma} &= 
\left(\nu_{\alpha\beta\gamma}^{g_2}\right) q_{\alpha\beta\gamma}
\left(g_2^* \nu_{\alpha\beta\gamma}^{g_1}\right) \cr
&\phantom{=}\ \times \left(\lambda_{\alpha\beta}^{g_1, g_2}\right)
\left(\lambda_{\beta\gamma}^{g_1, g_2}\right)
\left(\lambda_{\gamma\alpha}^{g_1, g_2}\right)
\left(\mu_{\alpha\beta}^{g_1 g_2}\right)
\left(\mu_{\beta\gamma}^{g_1 g_2}\right)
\left(\mu_{\gamma\alpha}^{g_1 g_2}\right) . 
}$$
Comparing these equations, we read the following constraint 
\eqn\qconst{
\left(\mu_{\alpha\beta}^{g_2}\right)
\left(g_2^* \mu_{\alpha\beta}^{g_1}\right) = 
\left(\lambda_{\alpha\beta}^{g_1, g_2}\right)
\left(\mu_{\alpha\beta}^{g_1 g_2}\right)
\left(\theta_{\alpha}^{g_1, g_2}\right)^{-1}
\left(\theta_{\beta}^{g_1, g_2}\right) ,
}
where $\theta_{\alpha}^{g_1,g_2}$ are some functions.
We also compute the pullbacks of $p^{\alpha\beta}$ by $g_1g_2$, 
$$\eqalign{
g_2^* (g_1^* p^{\alpha\beta}) &= p^{\alpha\beta} 
- d \log \mu_{\alpha\beta}^{g_2} + \Lambda^{(1)}(g_2)^{\alpha\beta}
+ \Lambda^{(1)}(g_2)^{\alpha} - \Lambda^{(1)}(g_2)^{\beta} \cr
&\phantom{=}\ - g_2^* \left(d \log \mu_{\alpha\beta}^{g_1}\right) 
+ g_2^* \Lambda^{(1)}(g_1)^{\alpha\beta} + g_2^* \Lambda^{(1)}(g_1)^{\alpha} 
- g_2^* \Lambda^{(1)}(g_1)^{\beta} , \cr
(g_1 g_2)^* p^{\alpha\beta} &= p^{\alpha\beta} 
- d \log \mu_{\alpha\beta}^{g_1 g_2} + \Lambda^{(1)}(g_1 g_2)^{\alpha\beta} 
+ \Lambda^{(1)}(g_1 g_2)^{\alpha} - \Lambda^{(1)}(g_1 g_2)^{\beta} \cr
&= p^{\alpha\beta} 
- d \log \theta_{\alpha}^{g_1, g_2} + d \log  \theta_{\beta}^{g_1, g_2} 
- d \log \mu_{\alpha\beta}^{g_2} 
- g_2^*\left(d \log \mu_{\alpha\beta}^{g_1}\right) \cr
&\phantom{=}\ + \Lambda^{(1)}(g_2)^{\alpha\beta} 
+ g_2^* \Lambda^{(1)}(g_1)^{\alpha\beta} 
- \Lambda^{(3)}(g_1, g_2)^{\alpha} + \Lambda^{(3)}(g_1, g_2)^{\beta} \cr
&\phantom{=}\ + \Lambda^{(1)}(g_1 g_2)^{\alpha} 
- \Lambda^{(1)}(g_1 g_2)^{\beta} .
}$$
Since $g_2^* (g_1^* p^{\alpha\beta})$ is equal 
to $(g_1 g_2)^* p^{\alpha\beta}$, 
we obtain the constraint 
\eqn\pconst{
\Lambda^{(1)}(g_2)^{\alpha} + g_2^* \Lambda^{(1)}(g_1)^{\alpha} = 
\Lambda^{(1)}(g_1 g_2)^{\alpha} - \Lambda^{(3)}(g_1, g_2)^{\alpha} 
- d \log \theta_{\alpha}^{g_1, g_2}
}
For $g_{\alpha\beta}$ we calculate the following pullbacks as
$$\eqalign{
g_2^*(g_1^* g_{\alpha\beta}) &= 
\left(g_2^* \mu_{\alpha\beta}^{g_1}\right)
\left(\mu_{\alpha\beta}^{g_2}\right) 
\left(g_2^* \rho_{\alpha}^{g_1}\right) 
\left(\rho_{\alpha}^{g_2}\right) g_{\alpha\beta} 
\left(\rho_{\beta}^{g_2}\right)^{-1} 
\left(g_2^* \rho_{\beta}^{g_1}\right)^{-1} \cr
&= \left(\lambda_{\alpha\beta}^{g_1,g_2}\right)
\left(\mu_{\alpha\beta}^{g_1g_2}\right)
\left(\theta_{\alpha}^{g_1,g_2}\right)^{-1}
\left(\theta_{\beta}^{g_1,g_2}\right) \cr
&\phantom{=}\ \times \left(g_2^* \rho_{\alpha}^{g_1}\right) 
\left(\rho_{\alpha}^{g_2}\right) g_{\alpha\beta} 
\left(\rho_{\beta}^{g_2}\right)^{-1} 
\left(g_2^* \rho_{\beta}^{g_1}\right)^{-1} , \cr
(g_1 g_2)^* g_{\alpha\beta} &= \left(\mu_{\alpha\beta}^{g_1 g_2}\right) 
\left(\rho_{\alpha}^{g_1 g_2}\right) g_{\alpha\beta}
\left(\rho_{\beta}^{g_1 g_2}\right)^{-1} .
}$$
From these equations, when 
$\lambda_{\alpha\beta}^{g_1,g_2}$ becomes one, we can obtain
\eqn\gconst{
\left(g_2^* \rho_{\alpha}^{g_1}\right)\left(\rho_{\alpha}^{g_2}\right) 
= \left(\theta_{\alpha}^{g_1, g_2}\right) 
\left(\rho_{\alpha}^{g_1 g_2}\right) .
}
We compare $\mu_{\alpha\beta}^{(g_1 g_2) g_3}$ 
with $\mu_{\alpha\beta}^{g_1 (g_2 g_3)}$. These two terms become 
$$\eqalign{
\mu_{\alpha\beta}^{(g_1 g_2) g_3} &= 
\left(\mu_{\alpha\beta}^{g_3}\right)\left(g_3^* \mu_{\alpha\beta}^{g_2}\right)
\left((g_2 g_3)^* \mu_{\alpha\beta}^{g_1}\right)
\left(\lambda_{\alpha\beta}^{g_1 g_2,g_3}\right)^{-1}
\left(g_3^* \lambda_{\alpha\beta}^{g_1,g_2}\right)^{-1} \cr
&\phantom{=}\ \times \left(\theta_{\alpha}^{g_1 g_2,g_3}\right) 
\left(g_3^* \theta_{\alpha}^{g_1,g_2}\right) 
\left(\theta_{\beta}^{g_1 g_2,g_3}\right)^{-1}
\left(g_3^* \theta_{\beta}^{g_1,g_2}\right)^{-1} , \cr
\mu_{\alpha\beta}^{g_1 (g_2 g_3)} &=
\left(\mu_{\alpha\beta}^{g_3}\right)\left(g_3^* \mu_{\alpha\beta}^{g_2}\right)
\left((g_2 g_3)^* \mu_{\alpha\beta}^{g_1}\right)
\left(\lambda_{\alpha\beta}^{g_1,g_2 g_3}\right)^{-1}
\left(\lambda_{\alpha\beta}^{g_2,g_3}\right)^{-1} \cr
&\phantom{=}\ \times \left(\theta_{\alpha}^{g_1,g_2 g_3}\right) 
\left(\theta_{\alpha}^{g_2,g_3}\right) 
\left(\theta_{\beta}^{g_1,g_2 g_3}\right)^{-1}
\left(\theta_{\beta}^{g_2,g_3}\right)^{-1} . 
}$$
From these equations and Eq.\gcix\ we obtain the constraint
\eqn\dtconst{
\left(\theta_{\alpha}^{g_1 g_2,g_3}\right)
\left(g_3^* \theta_{\alpha}^{g_1,g_2}\right) =
\gamma_{\alpha}^{g_1,g_2,g_3}
\left(\theta_{\alpha}^{g_1,g_2 g_3}\right)
\left(\theta_{\alpha}^{g_2,g_3}\right) . 
}
Note that we have required $\nu_{\alpha\beta\gamma}^g = 1$ and
$\lambda_{\alpha\beta}^{g_1,g_2} = 1$ and these conditions are 
consistent with the constraint \gcviii.

\newsec{Conclusions and discussion}

A lot of works on orbifolds and discrete torsion have been done in the string 
theories, but we do not know precisely these subjects in M-theory. 
So we considered orbifold and discrete torsion in M-theory 
on the analogy of the string theories.

We used the results shown in \SIV, 
where the three-form C-fields with connections were 
presented in terms of the three-forms $C^{\alpha}$, 
the two-forms $u^{\alpha\beta}$, 
the one-forms $v^{\alpha\beta\gamma}$ and 
the functions $h_{\alpha\beta\gamma\delta}$
defined on local patches. In \SIV\ 
the two-forms $\Lambda^{(2)}(g)^{\alpha}$, 
the one-forms $\Lambda^{(1)}(g)^{\alpha\beta}$ and 
the {\v C}ech cochains $\nu_{\alpha\beta\gamma}^g$ were also introduced as 
the structures describing the actions of orbifold group $\Gamma$ 
on the C-fields. For the constraints we introduced 
$\Lambda^{(3)}(g_1,g_2)^{\alpha}$, $\lambda^{g_1,g_2}_{\alpha\beta}$ and 
$\gamma_{\alpha}^{g_1,g_2,g_3}$, and 
the difference of orbifold actions $\omega^{g_1,g_2,g_3}$ was classified
by $H^3(\Gamma,U(1))$.

Firstly we studied the two-form fields $B$ in the worldvolume gauge theory 
on the branes. 
The two-form B-fields are linked to the three-form C-fields, 
because $C$ are transformed into $C + d B$ by the gauge transformations.
So we also described the B-fields and their connections on local patches by
the two-forms $B^{\alpha}$, the one-forms $p^{\alpha\beta}$ and 
the functions $q_{\alpha\beta\gamma}$. 
We wrote down the relation between 
the C-fields and the B-fields in Eqs.\bi, \bii\ and \biii.
From these equations and the orbifold group actions 
\gci, \gcii, \gciii\ and \gciv\ on the C-fields 
we obtained the actions on the B-fields, 
which are presented as Eqs.\gbiii, \gbii\ and \gbi, and then we introduced 
some {\v C}ech cochains $\mu_{\alpha\beta}^g$ and some local one-forms
$\Lambda^{(1)}(g)^{\alpha}$. These factors are the structures 
constructing the orbifold group actions on the B-fields. 
Calculating the pullbacks for the actions of $g_1g_2$ 
on $q_{\alpha\beta\gamma}$ 
and $p^{\alpha}$ in two different ways, we derived the 
constraints \qconst\ and \pconst, especially in Eq.\pconst\ we 
added some functions $\theta_{\alpha}^{g_1,g_2}$. 
In Eq.\gconst\ we are able to find that $\theta_{\alpha}^{g_1,g_2}$ 
plays a role similar to discrete torsion in the string theories.

Next we considered the one-form fields $A$. 
We also defined the A-fields on local 
patches in terms of the one-forms $A^{\alpha}$ 
and the matrices $g_{\alpha\beta}$. 
On the analogy of the string theories we described 
the relationships between the B-fields and the A-fields as Eqs.\ai\ and \aii. 
In order to define the orbifold group actions 
on $A^{\alpha}$ and $g_{\alpha\beta}$, we needed some conditions for 
$\nu_{\alpha\beta\gamma}^g$ and $\Lambda^{(1)}(g)^{\alpha\beta}$, 
which are the data for the orbifold group actions on the C-fields. 
The conditions are that 
$\nu_{\alpha\beta\gamma}^g$ becomes one and that 
$\Lambda^{(1)}(g)^{\alpha\beta}$ vanishes. 
They are satisfied when the C-fields are topologically trivial, 
that is, when 
$h_{\alpha\beta\gamma\delta} = 1$ and $v^{\alpha\beta\gamma} = 0$. 
Then we obtained the orbifold group actions \gaii\ on $g_{\alpha\beta}$ from 
Eqs.\gbiii\ and \aii, where we introduced some functions $\rho_{\alpha}^g$, 
and the actions \gai\ on $A^{\alpha}$ from 
Eqs.\gbii, \ai\ and \gaii.

Let us consider the specialized situation so that 
the C-fields are completely trivial, that is, $C^{\alpha}$ is constant, 
$u^{\alpha\beta}$ and $v^{\alpha\beta\gamma}$ vanish and 
$h_{\alpha\beta\gamma\delta}$ is equal to one. 
Since we can take $\nu_{\alpha\beta\gamma}^g = 1$ from Eq.\gciv, and 
$\lambda_{\alpha\beta}^{g_1,g_2} = 1$ from Eq.\gcviii, 
we replace $\gamma_{\alpha}^{g_1,g_2,g_3}$ with a globally defined constant
$\gamma^{g_1,g_2,g_3}$. 
We also require that the B-fields are completely trivial, in other words, 
$B^{\alpha}$ becomes globally constant, $p^{\alpha\beta}$ 
vanishes and $q_{\alpha\beta\gamma}$ is equal to one. 
We can take $\mu_{\alpha\beta}^g = 1$ from Eq.\gbiii\ and 
$\theta_{\alpha}^{g_1,g_2}$ are replaced with 
globally defined $\theta^{g_1,g_2}$ from Eq.\qconst. 
And we regard the A-fields as the topologically trivial fields. Since 
$g_{\alpha\beta}$ becomes one, from Eq.\gaii\ we obtain
$\rho_{\alpha}^g = \rho_{\beta}^g$. After all Eq.\gconst\ leads to
$$
\left(\rho^{g}\right) \left(\rho^{h}\right) = \left(\theta^{g,h}\right)
\left(\rho^{gh}\right), \quad g,h \in \Gamma.
$$
This equation implies that the representation of orbifold groups $\rho^g$ 
is projective. 
This result is similar to Eq.\d\ shown in 
\refs{\D,\DF}.

We were able to have more interesting features in M-theory. 
Calculating $\mu_{\alpha\beta}^{g_1g_2g_3}$ in two different ways, 
we obtained Eq.\dtconst. In the situation mentioned above 
we can define $\theta^{g_1,g_2}_{\alpha}$ and $\gamma^{g_1,g_2,g_3}_{\alpha}$
as global constants, and from Eq.\dtconst\ we obtain
$$
\left(\theta^{g_1 g_2,g_3}\right)
\left(\theta^{g_1,g_2}\right) = 
\gamma^{g_1,g_2,g_3}
\left(\theta^{g_1,g_2 g_3}\right)
\left(\theta^{g_2,g_3}\right) , \quad g_1, g_2, g_3 \in \Gamma .
$$
We should recall that $\omega^{g_1,g_2,g_3} 
(=\gamma^{g_1,g_2,g_3}/{\bar \gamma}^{g_1,g_2,g_3})$ 
are classified by $H^3(\Gamma,U(1))$. 
So we assume $\gamma^{g_1,g_2,g_3}$ 
as discrete torsion for the branes in M-theory.

In M-theory there exist membranes and M5-branes. 
Open membranes have end points and end lines on the M5-branes, and 
in a gauge theory on the worldvolume of $N$ coincident M5-branes we assume 
the end lines and the end points as two-form tensor fields 
and one-form gauge fields respectively. 
So we considered the two-form B-fields and the one-form A-fields 
in the three-form C-fields background. 
When the transverse space for the M5-branes is ${\bf R} \times {\bf C}^2$, 
the gauge theory on the worldvolume includes 
two complex scalars and one real scalars as fluctuations of the M5-branes
to the transverse directions. 
When the M5-branes are located at the singularity of orbifold 
${\bf R} \times {\bf C}^2 / \Gamma$, where ${\bf C}^2 / \Gamma$ is ALE space, 
the gauge theory becomes a 
six dimensional ${\cal N} = 1$ supersymmetric Yang-Mills theory 
for large $N$ \refs{\FKPZ,\AOT}. Then a tensor multiplet, 
a hypermultiplet and a vector multiplet consist 
of the two-form field and the real scalar field, of 
the complex scalars and of the one-form field respectively.
The moduli space of the Yang-Mills theory on the M5-branes at the 
orbifold singularity with discrete torsion may have good 
correspondences\SS\ to the geometric structure of orbifold 
on the analogy of \refs{\BLI,\BLII}.

Though we know that the end points of open strings have 
Chan-Paton factors in the string theories, 
the analogues of Chan-Paton factors as the end lines of open membranes 
are not clear in M-theory. 
So we do not know precisely what type of values 
$\theta^{g_1,g_2}$ and $\gamma^{g_1,g_2,g_3}$ are.
But at least we are able to suggest that the representation $\rho^g$ of 
the orbifold group is projective and that 
the phase $\theta^{g_1,g_2}$ has the structure 
which are determined by the M-theory discrete torsion $\gamma^{g_1,g_2,g_3}$.

We will need to make more precise mathematical analyses, 
for example, quotient stacks \refs{\SVIII,\SIX}\ and K-theory. 
Dp-branes and M-theory branes are studied in the contexts of 
K-theory \W\ and twisted K-theory \refs{\K,\DMW}. Since the p-branes are realized 
as (p+1)-form fields in low energy effective actions, there may exist 
some analogues of discrete torsion for the p-branes and the open p-branes.

\bigbreak\bigskip\bigskip\centerline{{\bf Acknowledgement}}\nobreak

I am grateful to S. Matsuda for encouragement.
This work is supported in part by JSPS Research Fellowships for Young 
Scientists (\#4783).

\listrefs

\bye